\documentclass[12pt]{article}
\usepackage{amsmath}
\usepackage{amssymb}
\usepackage{epsfig}
\usepackage{euscript}
\usepackage{fancybox}
\usepackage{color}

\def\mathswitchr#1{\relax\ifmmode{\mathrm{#1}}\else$\mathrm{#1}$\fi}

\newcommand {\pslash}{\hbox{$\not\hbox{\kern-2.3pt $p$}$}}

\usepackage{cite}

\usepackage{epic}

\def\alf1{ {\alpha\over\pi} }

\begin{document}
\begin{titlepage}
\begin{flushright}
{\bf BU-HEPP-08-12 }\\
{\bf Aug., 2008}\\
\end{flushright}
 
\begin{center}
{\Large Planck Scale Cosmology in Resummed Quantum Gravity$^{\dagger}$
}
\end{center}

\vspace{2mm}
\begin{center}
{\bf   B.F.L. Ward}\\
\vspace{2mm}
{\em Department of Physics,\\
 Baylor University, Waco, Texas, 76798-7316, USA}\\
\end{center}

\vspace{5mm}
\begin{center}
{\bf   Abstract}
\end{center}
We show that, by using resummation techniques
based on the extension of the methods of Yennie, Frautschi and Suura
to Feynman's formulation of Einstein's theory, we get quantum field theoretic
predictions for the UV fixed-point values of the dimensionless
gravitational and cosmological constants. Connections to
the phenomenological asymptotic safety analysis of Planck scale
cosmology by Bonanno and Reuter are discussed.
\vspace{10mm}
\vspace{10mm}
\renewcommand{\baselinestretch}{0.1}
\footnoterule
\noindent
{\footnotesize
\begin{itemize}
\item[${\dagger}$]
Work partly supported
by the US Department of Energy grant DE-FG02-05ER41399
and by NATO Grant PST.CLG.980342.
\end{itemize}
}

\end{titlepage}

\def\Kmax{K_{\rm max}}\def\ieps{{i\epsilon}}\def\rQCD{{\rm QCD}}
\renewcommand{\theequation}{\arabic{equation}}
\font\fortssbx=cmssbx10 scaled \magstep2
\renewcommand\thepage{}
\parskip.1truein\parindent=20pt\pagenumbering{arabic}\par
While the successes of the inflationary model~\cite{guth,linde} of cosmology
are well-known, there remains the deeper question of the origin
of the special scalar (inflaton) field required for its realization.
It opens the discussion for the possible fundamental dynamical mechanism
that may lead to the same realization and, thereby, provide a deeper
insight into the very origin of our Universe as we know it today.
In Ref.~\cite{reuter1,reuter2}, it has been argued that the phenomenological
asymptotic safety approach~\cite{laut,reuter3,litim,perc} to quantum gravity
may indeed provide such a realization: the attendant UV fixed point solution
allows one to develop Planck scale cosmology that joins smoothly onto
the standard Friedmann-Walker-Robertson classical descriptions so
that then one arrives at a quantum mechanical 
solution to the horizon, flatness, entropy
and scale free spectrum problems. Here, we show that in the new
resummed theory~\cite{bw1,bw2} of quantum gravity, 
we recover the properties as used in Refs.~\cite{reuter1,reuter2} 
for the UV fixed point of quantum gravity with the
added results that we get predictions for the fixed point values of
the respective dimensionless gravitational and cosmological constants
in their analysis. \par
Let us recapitulate the Planck scale cosmology presented phenomenologically
in Refs.~\cite{reuter1,reuter2}. The starting point is the Einstein-Hilbert 
theory
\begin{equation}
{\cal L}(x) = \frac{1}{2\kappa^2}\sqrt{-g}\left( R -2\Lambda\right)
\label{lgwrld1a}
\end{equation} 
where $R$ is the curvature scalar, $g$ is the determinant of the metric
of space-time $g_{\mu\nu}$, $\Lambda$ is the cosmological
constant and $\kappa=\sqrt{8\pi G_N}$ for Newton's constant
$G_N$. Using the phenomenological exact renormalization group
for the Wilsonian coarse grained effective 
average action in field space, the authors in Ref.~\cite{reuter1,reuter2} 
have argued that
the attendant running Newton constant $G_N(k)$ and running 
cosmological constant
$\Lambda(k)$ approach UV fixed points as $k$ goes to infinity
in the deep Euclidean regime in the sense that 
$k^2G_N(k)\rightarrow g_*,\; \Lambda(k)\rightarrow \lambda_*k^2$
for $k\rightarrow \infty$ in the Euclidean regime.\par
The contact with cosmology then proceeds as follows. Using a phenomenological
connection between the momentum scale $k$ characterizing the coarseness
of the Wilsonian graininess of the average effective action and the
cosmological time $t$, the authors
in Refs.~\cite{reuter1,reuter2} show that the standard cosmological
equations admit of the following extension:
\begin{align}
(\frac{\dot{a}}{a})^2+\frac{K}{a^2}&=\frac{1}{3}\Lambda+\frac{8\pi}{3}G_N\rho\\
\dot{\rho}+3(1+\omega)\frac{\dot{a}}{a}\rho&=0\\
\dot{\Lambda}+8\pi\rho\dot{G_N}&=0\\
G_N(t)&=G_N(k(t))\\
\Lambda(t)&=\Lambda(k(t))
\label{coseqn1}
\end{align}
in a standard notation for the density $\rho$ and scale factor $a(t)$
with the Robertson-Walker metric representation as
\begin{equation}
ds^2=dt^2-a(t)^2\left(\frac{dr^2}{1-Kr^2}+r^2(d\theta^2+\sin^2\theta d\phi^2)\right)
\label{metric1}
\end{equation}
so that $K=0,1,-1$ correspond respectively to flat, spherical and
pseudo-spherical 3-spaces for constant time t.  Here, the equation of state
is taken as a linear relation between the pressure $p$ and $\rho$,
\begin{equation} 
p(t)=\omega \rho(t),
\end{equation}
and the functional relationship between the respective
momentum scale k and the cosmological time t is determined
in Refs.~\cite{reuter1,reuter2} phenomenologically via
\begin{equation}
k(t)=\frac{\xi}{t}
\end{equation}
for some positive constant $\xi$ which then must be determined
from requirements on
physically observable predictions.\par
Using the UV fixed points as discussed above for $k^2G_N(k)$ and
$\Lambda(k)/k^2$ obtained from their phenomenological, exact renormalization
group (asymptotic safety) 
analysis, the authors in Refs.~\cite{reuter1,reuter2}
show that the system in (\ref{coseqn1}) admits, for $K=0$,
a solution in the Planck regime where $0\le t\le t_{\text{class}}$, with
$t_{\text{class}}$ a few times the Planck time $t_{Pl}$, which joins
smoothly onto a solution in the classical regime, $t>t_{\text{class}}$,
which coincides with standard Friedmann-Robertson-Walker phenomenology
but with the horizon, flatness, scale free Harrison-Zeldovich spectrum,
and entropy problems all solved by purely Planck scale quantum physics.\par
The phenomenological nature of the analysis is manifested in that
the fixed-point results $g_*,\lambda_*$ depend on the cut-offs
used in the Wilsonian coarse-graining procedure, for example. 
The key properties of $g_*,\lambda_*$ used for the analyses of Refs.~\cite{reuter1,reuter2} are that they are both positive and that the product 
$g_*\lambda_*$ is cut-off/threshold function independent.
Here, we present the predictions for these
UV limits as implied by resummed quantum gravity theory as presented in
~\cite{bw1,bw2}. In this way, we put the arguments in Refs.~\cite{reuter1,reuter2} on a more rigorous theoretical basis.\par
\begin{figure}
\begin{center}
\epsfig{file=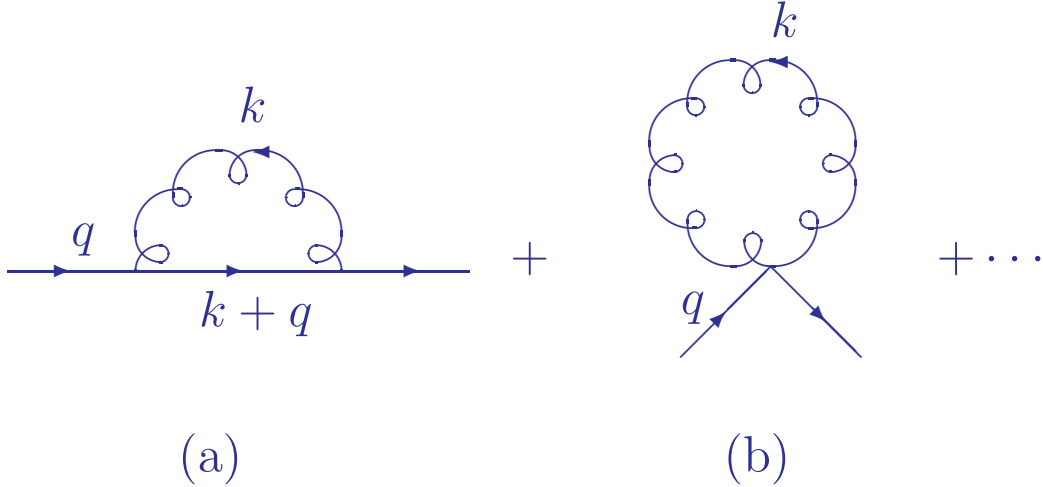,width=140mm}
\end{center}
\caption{\baselineskip=7mm     Graviton loop contributions to the
scalar propagator. $q$ is the 4-momentum of the scalar.}
\label{fig1}
\end{figure}

 
We start with the prediction for $g_*$, which we already presented in Refs.~\cite{bw1,bw2}. As the theory we use is not very familiar, we recapitulate
the main steps in the calculation so that our discussion is self-contained.
Referring to Fig.~\ref{fig1}, 
we have shown in Refs.~\cite{bw1,bw2} that the large virtual IR effects
in the respective loop integrals for 
the scalar propagator in quantum general relativity 
can be resummed to the {\em exact} result
\begin{equation}
i\Delta'_F(k)|_{\text{resummed}} =  \frac{ie^{B''_g(k)}}{(k^2-m^2-\Sigma'_s+i\epsilon)}
\label{resum}
\end{equation}
for{\small ~~~($\Delta =k^2 - m^2$)
\begin{equation}
\begin{split} 
B''_g(k)&= -2i\kappa^2k^4\frac{\int d^4\ell}{16\pi^4}\frac{1}{\ell^2-\lambda^2+i\epsilon}\\
&\qquad\frac{1}{(\ell^2+2\ell k+\Delta +i\epsilon)^2}\\
&=\frac{\kappa^2|k^2|}{8\pi^2}\ln\left(\frac{m^2}{m^2+|k^2|}\right),       
\end{split}
\label{yfs1} 
\end{equation}}
where the latter form holds for the UV regime, so that (\ref{resum}) 
falls faster than any power of $|k^2|$. An analogous result~\cite{bw1} holds
for m=0. As $\Sigma'_s$ starts in ${\cal O}(\kappa^2)$,
we may drop it in calculating one-loop effects. It follows that,
when the respective analogs of (\ref{resum}) are used for the
elementary particles, one-loop 
corrections are finite. It can be shown actually that the use of
our resummed propagators renders all quantum 
gravity loops UV finite~\cite{bw1,bw2}. We have called this representation
of the quantum theory of general relativity resummed quantum gravity (RQG).
\par
When we use our resummed propagator results, 
as extended to all the particles
in the SM Lagrangian and to the graviton itself, working now with the
complete theory
\begin{equation}
{\cal L}(x) = \frac{1}{2\kappa^2}\sqrt{-g} \left(R-2\Lambda\right)
            + \sqrt{-g} L^{\cal G}_{SM}(x)
\label{lgwrld1}
\end{equation}
where $L^{\cal G}_{SM}(x)$ is SM Lagrangian written in diffeomorphism
invariant form as explained in Refs.~\cite{bw1,bw2}, we show in Refs.~\cite{bw1,bw2} that the denominator for the propagation of transverse-traceless
modes of the graviton becomes
\begin{equation}
q^2+\Sigma^T(q^2)+i\epsilon\cong q^2-q^4\frac{c_{2,eff}}{360\pi M_{Pl}^2},
\label{dengrvprp}
\end{equation}
where we have defined
\begin{equation}
\begin{split}
c_{2,eff}&=\sum_{\text{SM particles j}}n_jI_2(\lambda_c(j))\\
         &\cong 2.56\times 10^4
\end{split}
\label{c2eff}
\end{equation}
with $I_2$ defined~\cite{bw1,bw2}
by
\begin{equation}
I_2(\lambda_c) =\int^{\infty}_0dx x^3(1+x)^{-4-\lambda_c x}
\end{equation}
and with $\lambda_c(j)=\frac{2m_j^2}{\pi M_{Pl}^2}$ and~\cite{bw1,bw2}
$n_j$ equal to the number of effective degrees of particle $j$. In arriving at (\ref{c2eff}), we take the SM
masses as follows: for the now presumed three massive neutrinos~\cite{neut},
we estimate a mass at $\sim 3$ eV; for
the remaining members
of the known three generations of Dirac fermions
$\{e,\mu,\tau,u,d,s,c,b,t\}$, we use~\cite{pdg2002,pdg2004}
$m_e\cong 0.51$ MeV, $m_\mu \cong 0.106$ GeV, $m_\tau \cong 1.78$ GeV,
$m_u \cong 5.1$ MeV, $m_d \cong 8.9$ MeV, $m_s \cong 0.17$ GeV,
$m_c \cong 1.3$ GeV, $m_b \cong 4.5$ GeV and $m_t \cong 174$ GeV and for
the massive vector bosons $W^{\pm},~Z$ we use the masses
$M_W\cong 80.4$ GeV,~$M_Z\cong 91.19$ GeV, respectively.
We set the Higgs mass at $m_H\cong 120$GeV, in view of the
limit from LEP2~\cite{lewwg}.
We note that (see the Appendix 1 in Ref.~\cite{bw1}) when the
rest mass of particle $j$ is zero, such as it is for the photon and the gluon,
the value of $m_j$ turns-out to be
$\sqrt{2}$ times the gravitational infrared cut-off
mass~\cite{cosm1}, which is $m_g\cong 3.1\times 10^{-33}$eV.
We further note that, from the
exact one-loop analysis of Ref.\cite{tHvelt1}, it also follows
that the value of $n_j$ for the graviton and its attendant ghost is $42$.
For $\lambda_c\rightarrow 0$, we have found the approximate representation
\begin{equation}
I_2(\lambda_c)\cong \ln\frac{1}{\lambda_c}-\ln\ln\frac{1}{\lambda_c}-\frac{\ln\ln\frac{1}{\lambda_c}}{\ln\frac{1}{\lambda_c}-\ln\ln\frac{1}{\lambda_c}}-\frac{11}{6}.
\end{equation} 
These results allow us to identify (we use $G_N$ for $G_N(0)$) 
\begin{equation}
G_N(k)=G_N/(1+\frac{c_{2,eff}k^2}{360\pi M_{Pl}^2})
\end{equation}
and to compute the UV limit $g_*$ as
\begin{equation}
g_*=\lim_{k^2\rightarrow \infty}k^2G_N(k^2)=\frac{360\pi}{c_{2,eff}}\cong 0.0442.
\end{equation}
We stress that this result has no threshold/cut-off effects in it.
It is a pure property of the known world.\par
Turning now to the prediction for $\lambda_*$, we use the Euler-Lagrange
equations to get Einstein's equation as 
\begin{equation}
G_{\mu\nu}+\Lambda g_{\mu\nu}=-\kappa^2 T_{\mu\nu}
\label{eineq1}
\end{equation}
in a standard notation where $G_{\mu\nu}=R_{\mu\nu}-\frac{1}{2}Rg_{\mu\nu}$,
$R_{\mu\nu}$ is the contracted Riemann tensor, and
$T_{\mu\nu}$ is the energy-momentum tensor. Working then with
the representation $g_{\mu\nu}=\eta_{\mu\nu}+2\kappa h_{\mu\nu}$
for the flat Minkowski metric $\eta_{\mu\nu}=\text{diag}(1,-1,-1,-1)$
we see that to isolate $\Lambda$ in Einstein's 
equation (\ref{eineq1}) we may evaluate
its VEV(vacuum expectation value of both sides). 
For any bosonic quantum field $\varphi$ we use
the point-splitting definition (here, :~~: denotes normal ordering as usual)
\begin{equation}
\begin{split}
\varphi(0)\varphi(0)&=\lim_{\epsilon\rightarrow 0}\varphi(\epsilon)\varphi(0)\cr
&=\lim_{\epsilon\rightarrow 0} T(\varphi(\epsilon)\varphi(0))\cr
&=\lim_{\epsilon\rightarrow 0}\{ :(\varphi(\epsilon)\varphi(0)): + <0|T(\varphi(\epsilon)\varphi(0))|0>\}\cr
\end{split}
\end{equation}
where the limit $\epsilon\equiv(\epsilon,\vec{0})\rightarrow (0,0,0,0)\equiv 0$
is taken from a time-like direction respectively. Thus, 
a scalar makes the contribution to $\Lambda$ given by
\begin{equation}
\begin{split}
\Lambda_s&=-8\pi G_N\frac{\int d^4k}{2(2\pi)^4}\frac{(2\vec{k}^2+2m^2)e^{-\lambda_c(k^2/(2m^2))\ln(k^2/m^2+1)}}{k^2+m^2}\cr
&\cong -8\pi G_N[\frac{3}{G_N^{2}64\rho^2}],\cr
\end{split}
\end{equation} 
where $\rho=\ln\frac{2}{\lambda_c}$ and we have used the calculus
of Refs.~\cite{bw1,bw2}. The standard equal-time (anti-)commutation 
relations algebra realizations
then show that a Dirac fermion contributes $-4$ times $\Lambda_s$ to
$\Lambda$. The deep UV limit of $\Lambda$ then becomes, allowing $G_N(k)$
to run as we calculated,
\begin{equation}
\begin{split}
\Lambda(k) &\operatornamewithlimits{\longrightarrow}_{k^2\rightarrow \infty} k^2\lambda_*,\cr
\lambda_*&=-\frac{c_{2,eff}}{960}\sum_{j}(-1)^{F_j}n_j/\rho_j^2\cr
&\cong 0.232
\end{split}
\end{equation} 
where $F_j$ is the fermion number of $j$, $n_j$ is the effective
number of degrees of freedom of $j$ and $\rho_j=\rho(\lambda_c(m_j))$.
We see again that $\lambda_*$ is free of threshold/cut-off effects. It is
a pure prediction of our world as we know it.
In an exactly supersymmetric theory, $\lambda_*$ would vanish.\par
For reference, the UV fixed-point calculated here, 
$(g_*,\lambda_*)\cong (0.0442,0.232)$, can be compared with the estimates
in Refs.~\cite{reuter1,reuter2}, 
which give $(g_*,\lambda_*)\approx (0.27,0.36)$, with the understanding
that the analysis in Refs.~\cite{reuter1,reuter2} did not include
the specific SM matter action and that there is definitely cut-off function
sensitivity to the results in the latter analyses. What we do see
is that the qualitative results that $g_*$ and $\lambda_*$ are 
both positive and are significantly less than 1 in size with $\lambda_*>g_*$
are true of our results as well.\par 
To sum up, we have put 
Planck scale cosmology~\cite{reuter1,reuter2} on a
more rigorous basis. We look forward to possible checks from experiment,
to which we return elsewhere~\cite{elsewh}.\par
\section*{Acknowledgments}
We thank Profs. L. Alvarez-Gaume and W. Hollik for the support and kind
hospitality of the CERN TH Division and the Werner-Heisenberg-Institut, MPI, Munich, respectively, where a part of this work was done.

\end{document}